\documentclass[prd,12pt,aps,showpacs,superscriptaddress,preprintnumbers,tightenlines,floatfix,nofootinbib,amssymb]{revtex4}
\usepackage{epsfig}

\newcommand{\beqn}{\begin{eqnarray}}
\newcommand{\eeqn}{\end{eqnarray}}
\newcommand{\eq}[1]{(\ref{#1})}
\newcommand{\cC}{{\cal C}}

\newcommand{\cF}{{\cal F}}
\newcommand{\cA}{{\cal A}}

\def\bbbone{{\mathchoice {\rm 1\mskip-4mu l} {\rm 1\mskip-4mu l}
{\rm 1\mskip-4.5mu l} {\rm 1\mskip-5mu l}}}

\begin{document}

\title{Kert\'esz line and embedded monopoles in QCD}

\author{M.N.~Chernodub}\affiliation{Institute of Theoretical and
Experimental Physics, B.Cheremushkinskaya 25, Moscow, 117259, Russia}

\preprint{ITEP-LAT/2005-10}

\begin{abstract}
We propose a new class of defects in QCD which can be viewed as
``embedded'' monopoles made of quark and gluon fields. These objects are explicitly
gauge-invariant and they closely resemble the Nambu monopoles in the
Standard Electroweak model. We argue that the ``embedded QCD monopoles'' are proliferating in
the quark gluon plasma phase while in the low-temperature hadronic phase
the spatial proliferation of these objects is suppressed. At realistic quark masses and
zero chemical potential the hadronic and quark-gluon phases are generally believed to be
connected by a smooth crossover across which all thermodynamic quantities are non-singular.
We argue that these QCD phases are separated by a well--defined boundary -- known
as the Kert\'esz line in condensed matter systems -- associated
with the onset of the proliferation of the embedded QCD monopoles in the quark gluon
plasma phase.
\end{abstract}


\date{September 5, 2005}

\maketitle

The phase diagram of Quantum Chromodynamics has a rich structure
in the ``chemical potential'' ($\mu$) -- ``temperature'' ($T$)
plane~\cite{ref:general:reviews:QCD}. In particular, at small
chemical potential QCD predicts an existence of a transition at
$T_c \approx 170$~MeV from the low-temperature hadronic (or,
``confinement'') phase to the high-temperature quark-gluon (or,
``deconfinement'') phase. It is generally believed that at
realistic quark masses this transition is a smooth crossover
across which all thermodynamic quantities and their derivatives
are non-singular~\cite{ref:general:reviews:QCD,ref:crossover:lattice:facts}.
This means that the traditional order parameters -- such as vacuum expectation
value (v.e.v.) of the Polyakov loop and the chiral condensate --
do not behave as order parameters of the QCD transition at small chemical potential.
At larger $\mu$ the phase transition re-emerges at a tricritical
point and then continues as the first-order phase transition.
At even higher chemical potential more exotic phases (such as the
color superconductor phase and the color-flavor locking phase)
appear~\cite{ref:general:reviews:QCD}. Below we
concentrate on the crossover region at moderately small chemical potential.

\begin{figure}[!thb]
\begin{center}
\begin{tabular}{cc}
\includegraphics[scale=0.64,clip=true]{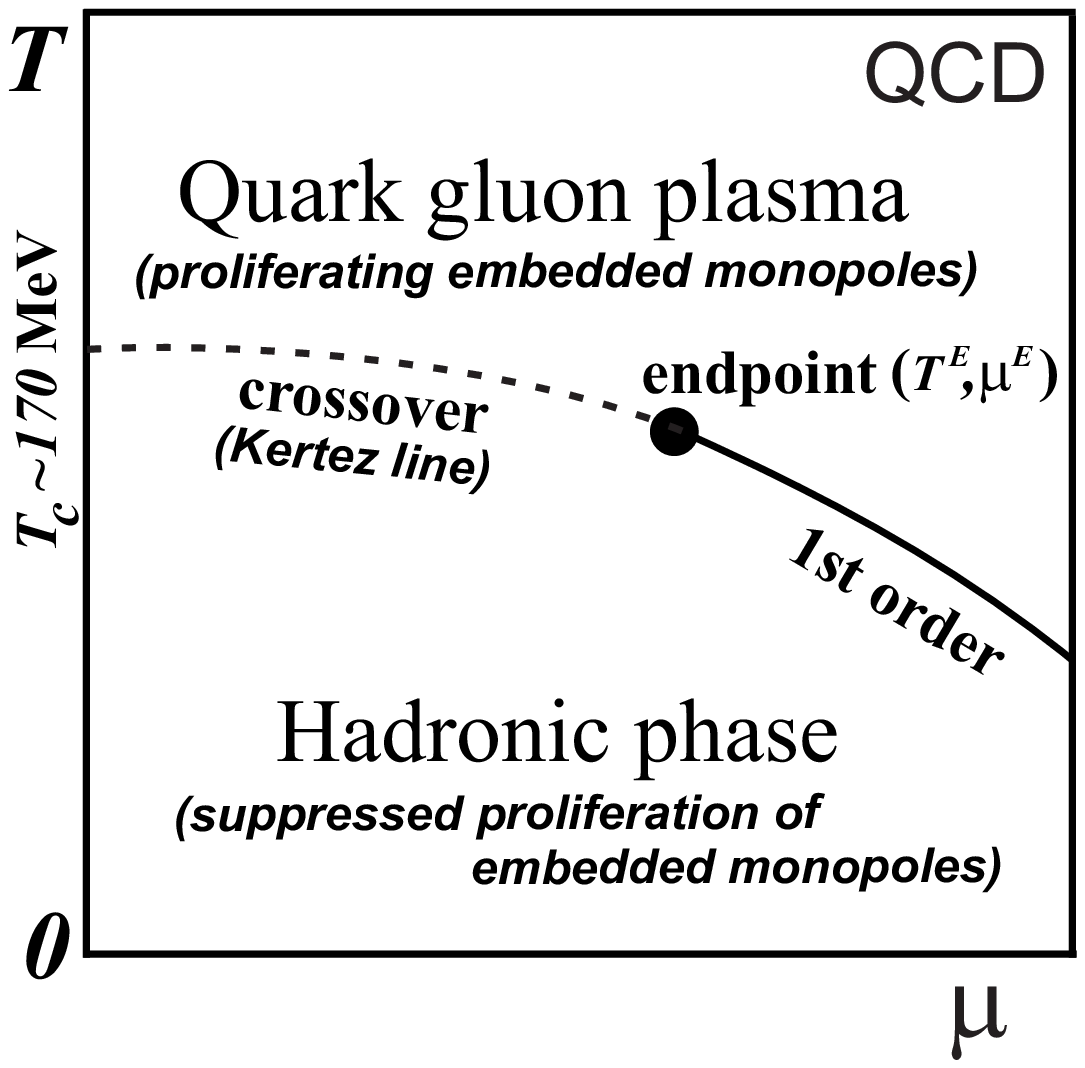} &
\hspace{8mm}
\includegraphics[scale=0.63,clip=true]{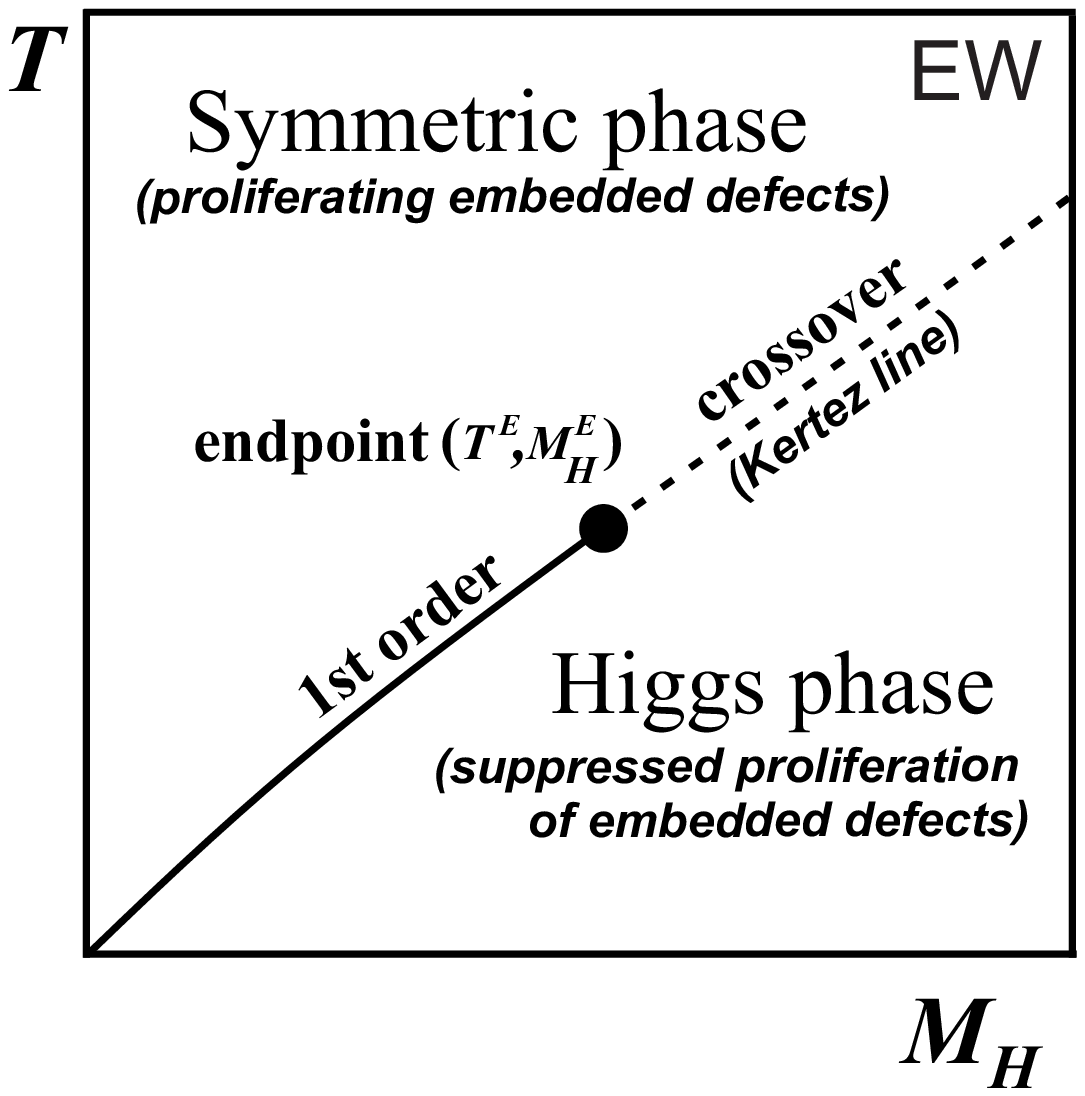} \\
(a) & (b)
\end{tabular}
\end{center}
\vspace{-4mm}
\caption{(a) QCD and (b) EW phase diagrams around tricritical points. The properties of the embedded
defects (suggested in QCD in this paper and found in the EW
model in Refs.~\cite{ref:chernodub:nambu,ref:hot:electroweak,ref:Z:percolation}) are indicated in the brackets. The tricritical
point, the first transition line and the Kert\'esz line are depicted as, respectively, the filled circle,
the solid line and the dashed line.}
\label{fig:phase}
\end{figure}

The $\mu$--$T$ phase diagram of QCD in a wide region around the tricritical point, Figure~\ref{fig:phase}(a),
is qualitatively similar to the phase diagram of the Standard model of Electroweak (EW)
interactions in the ``Higgs mass'' ($M_H$)--``temperature'' ($T$)
plane, Figure~\ref{fig:phase}(b). As it is well known, the symmetric (high-temperature) and
the Higgs (low-temperature) phases in the EW model are separated by a strong first order
EW phase transition at relatively small Higgs masses~\cite{ref:EW:phase}. As the Higgs mass
increases, the first order transition weakens and stops at a tricritical endpoint
$(T^E,M^{\mathrm{E}}_{H}) \approx (155\,\mathrm{GeV},72\,\mathrm{GeV})$
at which the transition is of the second order~\cite{ref:EW:phase,Rummukainen:1998as}.
At higher $M_H$ the phase transition becomes a smooth crossover across which
all thermodynamical quantities are smooth similarly to the case of QCD.

Another qualitative similarity between QCD and the EW model is
that both field theories do {\it not} possesses any topologically
stable monopole- or vortex-like defects. However, the absence of the
stable topological defects does not make the topological
structure of the EW model less interesting because it is well
known~\cite{ref:semilocal:review} that this model contains the
so-called ``embedded'' defects called the Nambu monopoles~\cite{ref:Nambu} and the
$Z$-vortices~\cite{ref:Z:vortex}.

Analytical arguments~\cite{ref:chernodub:nambu} as well as dynamical simulations of hot EW
model with the Higgs masses $M_H \sim 30$~GeV and $M_H \sim
70$~GeV show~\cite{ref:hot:electroweak} that the first order EW phase
transition is accompanied by the percolation transition of the
$Z$-vortices and the Nambu monopoles. These embedded defects are
suppressed in the Higgs phase and they are forming a dense
percolating (condensed) medium in the symmetric phase. As the mass
of the Higgs particle increases, the percolation transition does
not stop at the tricritical point and it continues into the
crossover region~\cite{ref:Z:percolation} still discriminating
between the high-- and low--temperature phases,
Figure~\ref{fig:phase}(b).

In the condensed matter physics, the percolation transition
realized in the absence of the thermodynamic phase transition is
usually referred to as the Kert\'esz line~\cite{ref:Kertesz}. The
simplest realization of the Kert\'esz line appears in the Ising
model in an external magnetic field. Each configurations of the
Ising spins can be associated with a set of the Fortuin--Kasteleyn
(FK) clusters~\cite{ref:Fortuin:Kasteleyn} which are defined as a
set of lattice links connecting nearest spins in the same spin
states. The FK clusters are known to be proliferating
(percolating) in the high temperature phase. As the temperature
gets lower the percolation of the FK clusters disappears
(in the absence of the external magnetic field) at the
phase transition (the Curie point). However, at non-zero external field
the partition function is analytic in temperature and the phase
transition is absent while the percolation transition (the
Kert\'esz line) still exists at any value of the external field.

The concept of the Kert\'esz line appears naturally in QCD without
reference to any (topological) defects. At high enough
temperature/density of the quark matter -- for example, in the
heavy-ion collision experiments -- the hadrons may overlap and
form clusters within which the quarks are no more confined. The
onset of the quark-gluon plasma phase may be associated with the
percolation transition of the hadron
clusters~\cite{ref:Satz:theory}. In the context of the field
theory the Kert\'esz line was also discussed for the
monopole~\cite{ref:Baig} and vortex~\cite{ref:Arwed} percolation
in compact U(1) Higgs models, for the Nambu
monopole~\cite{ref:chernodub:nambu}, the
$Z$-vortex~\cite{ref:hot:electroweak,ref:Z:percolation}, and the
center vortex~\cite{ref:Greensite:Faber} percolation in the case
of the SU(2) Higgs model.

In this paper we suggest that the finite-temperature crossover
transition in QCD can be considered as the Kert\'esz--type
transition associated with percolation of the ``quark embedded
defects'' made of the quarks and the gluons.

Consider the embedded topological defects in the EW model (for simplicity we consider
vanishing Weinberg angle, $\theta_W=0$).
The bosonic sector of this model is basically the SU(2) gauge model
with the Higgs doublet $\Phi(x) = {(\phi_1(x),\phi_2(x))}^T$.
Mathematically, the explicit definition of the embedded defects in the EW model
is based on the composite scalar field $\chi^a(x)$ constructed from the fundamental
Higgs field $\Phi$:
\beqn
\chi^a(x) = - \Phi^\dagger(x) \tau^a \Phi(x)\,,
\label{eq:chi}
\eeqn
where $\tau^a$ are the Pauli matrices acting in the isospin space.
The field $\chi^a$ transforms in the adjoint representation of gauge group
and can be treated similarly to the triplet Higgs field in the SO(3)
Georgi-Glashow model.

In the unitary gauge of the EW model, $\Phi(x) = {(0,\phi(x))}^T$,
the composite field $\chi$ gets automatically fixed to the SO(3) unitary gauge,
$\chi^a = |\vec \chi| \, \delta^{a3}$. The non--zero expectation value of the Higgs field
$\Phi$ in the Higgs (low-temperature) phase of the EW model
guarantees a non-zero expectation value of the composite field
$\chi$ because of the identity $\vec \chi^2 = (\Phi^\dagger \Phi)^2$. Thus, if the Higgs
field $\Phi$ resides near the classical minimum of the Higgs potential,
$\langle \Phi \rangle = {(0,\eta)}^T$, then the composite field $\chi$ does so near the
value\footnote{Acknowledging the qualitative nature of our work we neglect
quantum corrections to the expectation values of the composite operators.}
$\chi^a_0 \equiv \langle \chi^a \rangle = |\eta|^2\, \delta^{a3}$, or
\beqn
\langle \vec \chi^2 \rangle = \langle \vec \Phi^2 \rangle^2 \equiv |\eta|^4\,.
\label{eq:chi2:average}
\eeqn

The non-zero vacuum expectation value of the composite field $\chi^a$ in the Higgs phase makes it possible
to construct a monopole--like configuration of the EW fields -- called the electroweak
Nambu or the electroweak  monopole~\cite{ref:Nambu} -- in a manner
similar to the 't~Hooft--Polyakov~\cite{ref:thooft,ref:polyakov}
construction of a monopole in the Georgi-Glashow model. The position of the monopole singularity can be identified
with the help of the gauge invariant 't~Hooft tensor~\cite{ref:thooft},
\beqn
\cF_{\mu\nu}(\chi,W) = F^a_{\mu\nu}\, \hat{\chi}^a
+ \frac{1}{g} \epsilon^{abc} \hat{\chi}^a {(D^{\mathrm{ad}}_\mu \hat{\chi})}^b {(D^{\mathrm{ad}}_\mu \hat{\chi})}^c\,,
\qquad
\hat{\chi}^a  = \frac{\chi^a}{|\vec \chi|}\,,
\label{eq:thooft:tensor}
\eeqn
where $\hat{\chi}^a \equiv \hat{\chi}^a(\Phi)$ is the unit color vector,
pointing into the direction of the composite $\chi$-field~\eq{eq:chi},
$F^a_{\mu\nu} \equiv F^a_{\mu\nu}(W)$ is the field strength tensor for the SU(2) gauge field $W^a_\mu$,
and ${(D^{\mathrm{ad}}_\mu)}^{ab} = \delta^{ab}\, \partial_\mu + g \, \epsilon^{abc} W^c_\mu$
is the adjoint derivative. Equation~\eq{eq:thooft:tensor} defines the gauge-invariant field strength tensor for the
$Z$--component of the gauge field, $Z_\mu = W^a_\mu \hat{\chi}^a$.

The current of the Nambu monopole,
\beqn
k^{\mathrm{EW}}_\nu = \partial_\mu \tilde \cF_{\mu\nu} \equiv
\int_\cC \frac{\partial X^\cC_\nu(\tau)}{\partial \tau}\, \delta^{(4)}(x - X(\tau))\,,
\qquad
\tilde \cF_{\mu\nu} = \frac{1}{2} \epsilon_{\mu\nu\alpha\beta} \cF_{\mu\nu}\,,
\label{eq:k:Nambu}
\eeqn
has a $\delta$--like singularity at the monopole worldline $\cC$ parameterized by the vector $x_\mu = X^\cC_\mu(\tau)$.
The location of the embedded monopoles are encoded in the gauge fields $W_\mu$ and the Higgs fields~$\Phi$ via
relations~(\ref{eq:chi},\ref{eq:thooft:tensor},\ref{eq:k:Nambu}).

The integration of the $Z$-magnetic flux~\eq{eq:thooft:tensor} over an infinitesimally small sphere surrounding
the monopole singularity gives the $Z$-magnetic charge of the monopole, which is quantized
in units of the elementary monopole charge~\cite{ref:semilocal:review}, $g_m = 4 \pi /g$.
The $Z$-magnetic charge of the Nambu monopole is conserved by virtue of its
definition~\eq{eq:k:Nambu}, $\partial_\mu k^{\mathrm{EW}}_\mu = 0$, and therefore the
Nambu monopoles can only disappear by annihilating with
anti-monopoles~\cite{ref:semilocal:review}. Finally, one should mention that in the Unitary gauge,
$\hat{\chi}^a = \delta^{a3}$, the electroweak monopole is just an Abelian monopole singularity in the
diagonal gauge field $Z_\mu \equiv W^3_\mu$. Therefore the electroweak monopole is an Abelian
monopole ``embedded'' into the EW model.

In the Nambu construction, the $Z$--magnetic flux is coming to an
isolated electroweak monopole along a semi-infinite
$Z$--vortex~\cite{ref:Z:vortex}. The $Z$-vortex can be considered
as the Abrikosov-Nielsen-Olesen vortex
configuration~\cite{ref:ANO} of the Abelian Higgs model embedded
into the EW model.

Both the $Z$-vortices and the Nambu monopoles are not stable
objects~\cite{ref:semilocal:review}. The monopoles decay via
annihilation with anti-monopoles while the $Z$--string can also
decay into the vacuum via a production of the $W$-bosons. The
tension of the $Z$-vortices and mass of the Nambu monopoles are
proportional to the appropriate powers of the Higgs expectation
value, $\eta$. Therefore at low temperatures -- where the Higgs
field has a large expectation value -- the embedded defects are
drastically suppressed and their effect on the system properties
is negligible. However in the symmetric phase at high
temperatures, as we have mentioned earlier, the thermal
fluctuations create a dense percolating medium of the embedded
defects~\cite{ref:chernodub:nambu,ref:hot:electroweak,ref:Z:percolation}.

Coming closer to QCD, let consider for simplicity the SU(2) gauge theory with one species of a
(generally, massive) fermion field~$\psi$ which transforms in the fundamental representation of
the gauge group. Then one can construct two SU(2) QCD analogues of the
electroweak $\chi$--field~\eq{eq:chi}:
\beqn
\xi^a_\Gamma = \bar\psi(x) \Gamma \tau^a \psi(x)\,,\qquad \Gamma = \bbbone\,, \gamma_5\,,
\label{eq:xi}
\eeqn
where both the scalar $\xi^a$ and the pseudoscalar $\xi^a_5$ fields are the real-valued triplet
vectors in the isospin space (we drop the index $\Gamma$ in $\xi_\Gamma$ if $\Gamma = \bbbone$).

The existence of the isospin vectors~\eq{eq:xi} allows us to define
the currents of the gauge-invariant monopoles in the SU(2) QCD in
a manner similar to the EW construction~\eq{eq:k:Nambu}:
\beqn
k^\Gamma_\nu   = \partial_\mu \tilde \cF_{\mu\nu}(\xi_\Gamma,A)\,,
\label{eq:k}
\eeqn
where $\cF_{\mu\nu}$ is the 't~Hooft tensor~\eq{eq:thooft:tensor} in which the EW gauge field $W_\mu$ is
replaced by the SU(2) gluon field $A_\mu$, and the EW composite field $\chi$ is substituted
by the fermionic composite fields~\eq{eq:xi}. Equation~\eq{eq:k} provides an explicitly
gauge-independent way to identify monopole-like singularities in QCD using the fermionic degrees of
freedom along the ideological line of Ref.~\cite{ref:chernodub:zakharov}. The location of the embedded
QCD monopoles is encoded in the gluon $A^a_\mu$ and fermion~$\psi$ fields via
relations~(\ref{eq:xi},\ref{eq:thooft:tensor},\ref{eq:k}).

The $k_\mu$ and $k^5_\mu$ fermionic monopoles carry the magnetic
charge with respect to, correspondingly, $\cA_\mu = A^a_\mu
\hat{\xi}^a$ and $\cA^5_\mu = A^a_\mu \hat{\xi}_5^a$ components of
the gauge field. In the Unitary, $\hat{\xi}^a = \delta^{a3}$, (or,
in the ``pseudo-Unitary'', $\hat{\xi}^a_5 = \delta^{a3}$) gauge
the $k_\mu$ (or, respectively, $k^5_\mu$) fermionic monopoles
correspond to monopoles ``embedded'' into the diagonal component,
$A^3_\mu$, of the gluon field. One can also consider these
monopoles as the Abelian monopoles determined in an Abelian
gauge~\cite{ref:tHooft:projection} which is defined by a
requirement of diagonalization of the corresponding composite
fermionic field~\eq{eq:xi}. Finally, in gauges, in which the
gauge field $A_\mu$ is smooth (presumably, in the Landau gauge),
one can consider the embedded monopoles as the hedgehogs in
the composite quark fields \eq{eq:xi}.

Thus, in the toy case of the $N_f=1$ SU(2) gauge theory one can define two
types of the embedded QCD monopoles, the currents of which are vector and
pseudo-vector variables~\eq{eq:k}. The existence of the
topologically nontrivial monopoles~\eq{eq:k} is not a dynamically
motivated fact. Instead, it is a simple (kinematical) consequence
of the existence of the adjoint real-valued fields~\eq{eq:xi},
which are not required to be condensed~\cite{ref:wetterich}.

In the real case of QCD the zoo of the embedded monopoles is much reacher.
Indeed, in the SU(3) gauge theory with $N_f$ massive fermions one can
introduce two matrices in the flavor space instead of two composite
scalar fields~\eq{eq:xi}:
\beqn
\Xi^a_{ff',\Gamma}(x) = \bar\psi_{f}(x) \Gamma \lambda^a \psi_{f'}(x)\,,
\label{eq:xi:matrix}
\eeqn
where $\lambda^a$, $a=1,\dots,8$ are the SU(3) Gell-Mann color matrices and $f,f' = 1,\dots,N_f$ are the flavor indices.
Each element of these matrices transforms in the adjoint representation of the SU(3) gauge group.

To characterize the quark embedded defects in QCD we use the fact
that the global flavor symmetry is explicitly broken by mass terms
at the Lagrangian level (we consider the realistic case of
non-equal quark masses). Using flavor transformations one can
rotate the quark fields into a flavor basis where the mass matrix
is diagonal. In this basis the diagonal elements of the
matrices~\eq{eq:xi:matrix} should be considered as the real-valued
color octet fields $\xi^a_{f,\Gamma} \equiv \Xi^a_{ff,\Gamma}$ (no
summation over the index $f$). The diagonal elements
$\xi^a_{f,\Gamma}$ are then used to construct the gauge-invariant
embedded monopoles as in the toy $N_c=2$, $N_f=1$ case~\eq{eq:k}.
Given the octet vectors~\eq{eq:xi:matrix} the monopole charges in
the $N_c=3$ color case can be characterized by integer magnetic
charges similarly to the monopoles in the $SU(N_c)$ Higgs
models~\cite{ref:yasha}.

Thus in QCD with $N_f$ massive fermions there are two types of
monopoles associated with each quark field ($i.e.$, we have $2N_f$
embedded monopoles in total). The trajectories and charges of
these defects can be defined analogously Eq.~\eq{eq:k}. In
principle, one can also define the ``mixed'' defects which involve
quark-antiquark bilinears of different flavors.

The composite quark fields $\xi^a_f$ and $\xi^a_{f,5}$ play role
of the adjoint Higgs field in the SU(3) version of the
Georgi-Glashow model. The existence of the stable monopoles in the
Georgi-Glashow model is guaranteed by the spontaneous breaking of
the SU(3) symmetry by the Higgs condensate. Contrary to the
Georgi-Glashow model, the color symmetry in QCD is known to
be unbroken~\cite{ref:wetterich}. Nevertheless we argue below, that in QCD the role of
the Higgs condensate is played by chiral condensates~\footnote{For recent reviews on
condensates in QCD see Ref.~\cite{ref:condensates}.} which
make the definition of the embedded QCD monopoles physically
meaningful.

Let us discuss the dynamical properties of the embedded monopoles
at finite temperature in the physically interesting $N_f=2$ case
of the two light $u$ and $d$ quarks of equal masses. The properties of the
defects can be guessed from the behavior of the
condensates constructed from the octet field $\xi^a = \sum_f \Xi^a_{ff}$
and the axial octet field $\xi^a_5 = \sum_f \Xi^a_{ff,5}$.
Obviously, due to the unbroken color invariance the simplest condensates
vanish, $\langle\xi^a\rangle = \langle\xi^a_5\rangle = 0$. The strength
of the condensates is characterized by the v.e.v. of the squared of the octet
fields which are nothing but the four-quark condensates of the form
$\langle{\vec \xi}^2_\Gamma\rangle \equiv \langle \bar \psi \Gamma \lambda^a \psi \, \bar \psi \Gamma \lambda^a \psi\rangle$
with $\Gamma=1,\,\gamma_5$. The factorization hypothesis~\cite{ref:SVZ} makes it possible to express the
four-quark condensates in terms of the chiral condensate $\langle \bar \psi \psi\rangle$,
\beqn
\langle{\vec \xi}^2_\Gamma\rangle = C_\Gamma \, \langle \bar \psi \psi\rangle^2\,,
\label{eq:xi2:average}
\eeqn
where $C_\Gamma$ is a numerical factor, and $\psi = u$ or $d$.

The QCD relation~\eq{eq:xi2:average} is remarkably similar to the
EW relation~\eq{eq:chi2:average}. At low temperatures the
composite octet fields $\xi^a_\Gamma$ are large similarly to the
low-temperature behavior of the composite $\chi$--field in the EW
model. As temperature increases, both the four-quark condensate in
QCD and the Higgs expectation value in the EW model are diminishing,
and they to small but non-vanishing values at the
corresponding crossover temperatures.

In the EW model the large zero-temperature value of the
$\chi$--field condensate gives rise to a large mass of the Nambu
monopoles which suppresses the monopole formation. This
fact may be understood intuitively since the field $\chi$ must be
vanishing inside the core of the Nambu monopole and this is
unfavorable in the presence of the $\chi$-condensate. Similarly,
the embedded monopoles in QCD force the octet fields
$\xi^a_\Gamma$ to be vanishing in the center of the monopole in
order to support their hedgehog structure. This is energetically
unfavorable at low temperatures because of the presence of the
four-quark condensates~\eq{eq:xi2:average}. However, as the
temperature (and the chemical potential) increases, the condensates~\eq{eq:xi2:average} continuously
melt and the suppression of monopoles becomes less and less effective. At
very high temperatures the value of the condensates is negligibly
small and the embedded QCD monopoles must form a dense and
percolating network -- supported by thermal fluctuations --
similarly to the behavior of the embedded EW
defects~\cite{ref:hot:electroweak}. Since at realistic quark masses
the phase transition is absent~\footnote{Note that our considerations
remain valid in the case of infinitely heavy quarks corresponding
to the quenched approximation. In this case the onset of percolation
should happen at the phase transition line.}, the onset of the
percolation transition marks the Kert\'esz line in QCD as shown
in Figure~\ref{fig:phase}(a) by the dashed line. It seems very plausible
that the percolation of the hadron clusters in the quark gluon
phase~\cite{ref:Satz:theory} is related to the proliferation of the
embedded QCD monopoles.

Our considerations are based on the temperature behavior of the
four-quark condensates which is qualitative
valid~\cite{ref:Johnson:Kisslinger} beyond the simple
factorization formula~\eq{eq:xi2:average}. Moreover, quantitative
estimations of Ref.~\cite{ref:Johnson:Kisslinger} show that as
temperature increases the v.e.v of the pseudoscalar octet fields
is dropping faster compared to the octet fields. Therefore the QCD
Kert\'esz line, Figure~\ref{fig:phase}(a), may in fact be
split into two lines since the onset of the percolation of the
pseudoscalar embedded monopoles may happen at (much) lower
temperature compared to the monopoles associated with the scalar
$\xi$--field. In the real QCD case the Kert\'esz line should
inevitably be split because the onset of percolation of the
embedded monopoles associated with different quark fields should
happen at different temperatures due to the difference in the
quark masses.

Summarizing, we proposed a new class of defects in QCD, the
embedded monopoles, which are made of quark and gluon fields. We
provided arguments in favor of existence of the percolation
transition (the Kert\'esz line) at the crossover regime in the QCD
phase diagram. The Kert\'esz line is associated with the onset of
the proliferation of these defects in the quark gluon phase. At
low temperature the formation of the embedded monopoles is
suppressed due to the presence of the four-quark condensates.

\newpage

\begin{acknowledgments}
The author is supported by grants RFBR 04-02-16079, RFBR 05-02-16306, DFG 436 RUS 113/739/0 and MK-4019.2004.2.
The author is grateful to N.O.~Agasian, E.~Gubankova, F.V.~Gubarev, B.O.~Kerbikov, S.M.~Morozov
and M.I.~Polikarpov for interesting discussions.
\end{acknowledgments}

\end{document}